\documentstyle[prl,aps,multicol,epsf]{revtex}
\begin{document}
\draft
\title{Landau mapping and Fermi liquid parameters of the 2D $t$$-$$J$ model}
\author{S. Nishimoto$^1$, Y. Ohta$^1$, and R. Eder$^2$}
\address{$^1$Department of Physics, Chiba University, Inage-ku, 
Chiba 263, Japan\\
$^2$Department of Solid State Physics, University of Groningen, 
9747 AG Groningen, The Netherlands}
\date{\today}
\maketitle
\begin{abstract}
An exact-diagonalization technique on small clusters is used to study the 
momentum distribution function $n({\bf k})$ of the lightly doped $t$$-$$J$ 
model in two-dimension (2D).  We find that $n({\bf k})$ can be decomposed 
into two components with bosonic and fermionic doping dependence.  
The bosonic component originates from the incoherent motion of holes and 
has no significance for low-energy physics.  For the fermionic component 
we explicitly perform the one-to-one Landau mapping between all low-lying 
states of the $t$$-$$J$ model and those of a system of spin-1/2 quasiparticles 
and extract the quasiparticle dispersion and Landau parameters.  
These results demonstrate that the $t$$-$$J$ model is a Fermi liquid with a 
`small' Fermi surface and a moderate attractive interaction between the 
quasiparticles.
\end{abstract}
\pacs{74.20.-Z, 75.10.Jm, 75.50.Ee}
\begin{multicols}{2}

\begin{figure}
\epsfxsize=8cm
\vspace{-0.5cm}
\hspace{0.0cm}\epsffile{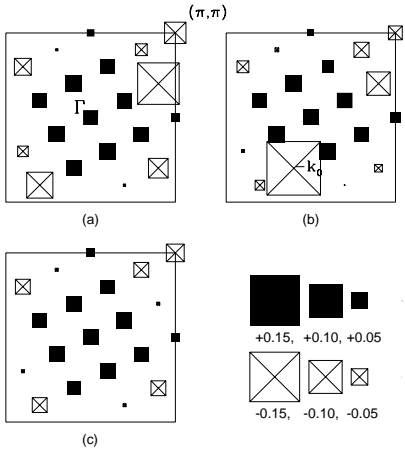}
\vspace{-1.5cm}
\narrowtext
\caption[]{EMD for the single-hole ground state of the $20$-site cluster 
(with total momentum ${\bf k}_0$$=$$(\pi/5,3\pi/5)$, $S_z=1/2$, and 
$J/t$$=$$0.5$).  The figures show (a) $n_\uparrow({\bf k})$, 
(b) $n_\downarrow({\bf k})$, and (c) $n_{\rm inc}({\bf k})$, in the 
entire Brillouin zone.  The edge of the square centered on a given 
${\bf k}$-point is proportional to $|n_{\sigma}({\bf k})-N_\sigma/N|$.  
Positive (negative) values are indicated by black (crossed) squares, 
and the `calibration' for their magnitude is given in the bottom 
right figure.  }
\label{fig1} 
\end{figure}

\begin{figure}
\epsfxsize=6cm
\vspace{-0.3cm}
\hspace{1.0cm}\epsffile{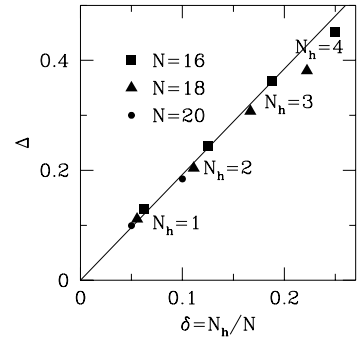}
\vspace{-1.25cm}
\narrowtext
\caption[]{Amplitude $\Delta$$=$$n(0,0)$$-$$n(\pi,\pi)$ as a function 
of the hole concentration $\delta$ evaluated for various cluster 
sizes $N$ and hole numbers $N_h$, where for odd $N_h$ the average over 
spin directions has been taken.  $J/t$$=$$0.4$ is used.  The full line 
is a guide to the eye.  }
\label{fig2} 
\end{figure}

\begin{figure}
\epsfxsize=10cm
\vspace{-0.5cm}
\hspace{-1.0cm}\epsffile{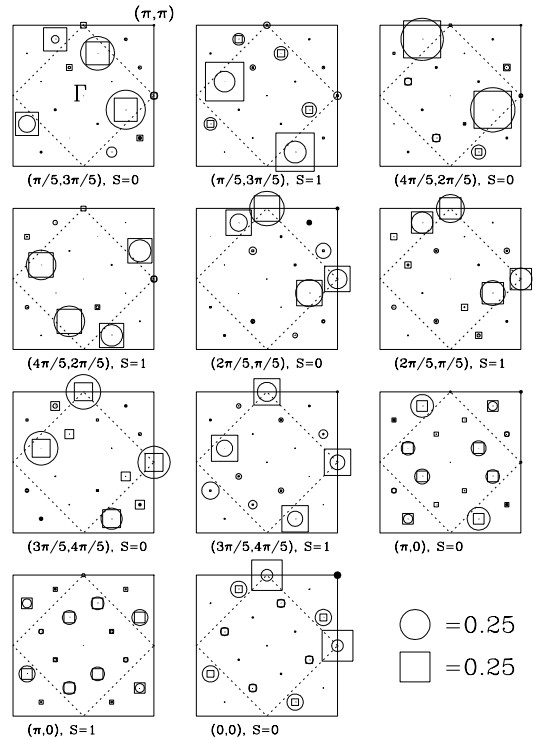}
\vspace{0.2cm}
\narrowtext
\caption[]{Comparison between the `coherent component' $n_{\rm coh}({\bf k})$ 
for all exact low-energy states of the $20$-site cluster $t$$-$$J$ model 
with $2$ holes (circles) at $J/t$$=$$0.5$ and the EMD of the quasihole 
Hamiltonian Eq.~(\ref{heff}) with $2$ particles (squares).  Each graph 
is labeled by the total momentum and spin of the respective eigenstate 
and shows the entire Brillouin zone.  The diameters of the circles 
(edges of the squares) centered on each ${\bf k}$-point are proportional 
to the respective $n({\bf k})$ (see the lower right edge for the `gauge').  
Momenta for the quasihole Hamiltonian are $-$$1$ times those for the $t$$-$$J$ 
model.}
\label{fig3} 
\end{figure}

\begin{figure}
\epsfxsize=10cm
\vspace{-0.5cm}
\hspace{-1.0cm}\epsffile{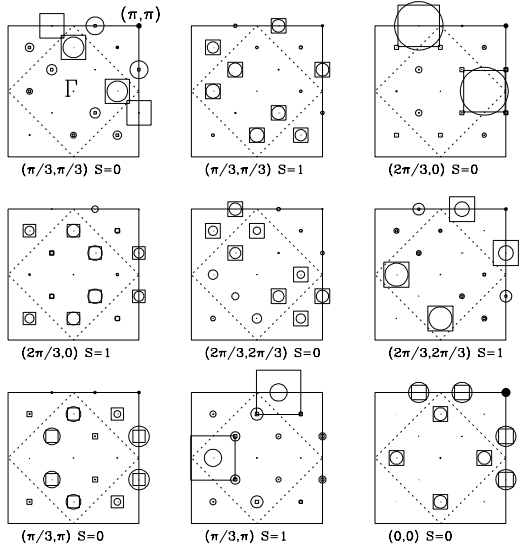}
\vspace{-2.0cm}
\narrowtext
\caption[]{Same as Fig.~\ref{fig3} but for $2$ holes in the $18$-site 
cluster.}
\label{fig4} 
\end{figure}

\begin{figure}
\epsfxsize=8cm
\vspace{-0.25cm}
\hspace{0.5cm}\epsffile{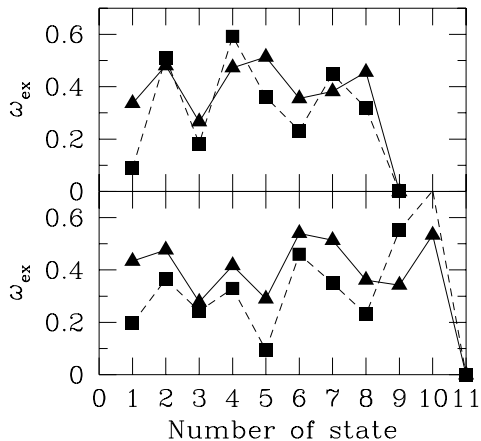}
\vspace{-1.75cm}
\narrowtext
\caption[]{Excitation energies $\omega_{\rm ex}$ for the eigenstates states
shown in Figs.~\ref{fig3} and \ref{fig4}: triangles are for the $t$$-$$J$ 
model and squares are for the quasiparticle Hamiltonian Eq.~(\ref{heff}).  
The numbers in the horizontal axis indicate the panels in 
Figs.~\ref{fig3} and \ref{fig4}, counted from top left, right way, 
to bottom right.}
\label{fig5} 
\end{figure}

Despite considerable experimental efforts, the Fermi-surface topology of 
cuprate superconductors continues to pose an intriguing and not really 
well-understood problem.  While it seemed to be settled for some time that 
the Fermi surface of these materials is simply the one predicted by the 
local-density approximation (LDA) with an only moderate `correlation narrowing' 
of the bandwidth, recent developments in photoelectron spectroscopy, like the 
discovery of the `shadow bands'\cite{Aebi} or the temperature dependent 
pseudogap at $(\pi,0)$\cite{Loeser}, have challenged this point of view.  
There is moreover the long-standing problem that Fermi-liquid--like 
calculations based on the LDA Fermi-surface cannot describe the doping 
dependence of either Hall constant or dc-resistivity on the hole 
concentration $\delta$; both quantities consistently suggest a carrier 
density $\propto \delta$\cite{Batlogg}, rather than $\propto (1-\delta)$ 
as it would be for the Fermi surfaces predicted by the LDA.  
On the other hand, taking the shadow bands as true part of the Fermi 
surface, which then would have the topology of elliptical hole pockets 
centered on $(\pi/2,\pi/2)$, would immediately lead to complete accord 
between Fermi-surface topology and transport properties in the framework 
of a very simple Fermi-liquid--like picture\cite{Trugman}.  While such a 
picture thus is quite appealing, clear experimental evidence for hole 
pockets has not been found so far\cite{Ding}.\\ 
It is the purpose of the present paper to address the problem of 
Fermi-surface topology of cuprates theoretically, by studying the 
2D $t$$-$$J$ model, the simplest strong correlation model which may give 
a realistic description of the CuO$_2$ plane.  To that end we analyze 
the electron momentum distribution (EMD) function 
$n_{\sigma}({\bf k})$$=$$\langle c_{{\bf k}\sigma}^\dagger c_{{\bf k}\sigma}
\rangle$.  In both the conventional Fermi liquid, as well as more exotic 
quantum liquids such as the Luttinger liquid, the EMD shows singularities 
which mark the location in ${\bf k}$-space of the extreme low-energy 
single-particle--like excitations.  
While $n({\bf k})$ (for simplicity we omit the spin index unless it is 
indispensable) can be readily evaluated exactly for small clusters of 
strong correlation models, one may not necessarily gain much from such data.  
The reason is that in the photoemission spectra of strongly correlated 
electrons the larger part of spectral weight resides in incoherent continua 
whereas the quasiparticle peak near the Fermi energy $E_{\rm F}$ carries 
only a small fraction of the weight.  Since $n({\bf k})$ is the zeroth 
moment of the spectral weight, it can be decomposed as a sum of the 
quasiparticle part, which is responsible for the step of magnitude $Z$ at 
the Fermi surface, and an incoherent background: 
\begin{equation}
n({\bf k}) = Z\cdot \Theta(\omega_{\rm QP} - E_{\rm F}) + \int\!\! 
{\rm d}\omega\, A_{\rm inc}({\bf k},\omega),
\end{equation}
where $\omega_{\rm QP}$ denotes the quasiparticle dispersion and 
$A_{\rm inc}({\bf k},\omega)$ the incoherent high-energy part of the spectral 
function.  In general the latter part has a ${\bf k}$-dependence of its 
own which is superimposed over the true Fermi surface discontinuities.  
In principle, the incoherent background should be a smooth function of 
${\bf k}$, so that a sufficiently good momentum resolution would allow 
to distinguish the step-like variations at $E_{\rm F}$; this however 
is out of the question with the relatively coarse ${\bf k}$-meshes 
available in cluster diagonalization, so that any simple assignment of 
a Fermi surface is impossible (see Ref.~\cite{comment} for a counterexample).  
On the other hand, numerical diagonalization yields the exact EMD for 
in principle {\it all} low-energy states without any statistical error, 
and in the following we want to take advantage of this feature.  
We will show that the smoothly varying incoherent component of $n({\bf k})$ 
is practically identical for all the low-energy states with given hole number.  
This allows to subtract off this incoherent component from the EMD for 
the various low-energy states, and thus make visible the `coherent structures' 
in $n({\bf k})$, which originate from the true Fermi-surface physics.  
These coherent structures, which give the ${\bf k}$-space distribution of 
the {\it quasiparticles} rather than the bare electrons, then can be used 
to establish the Landau mapping between the exact low-energy eigenstates 
of the $t$$-$$J$ model to those of a suitably chosen quasiparticle 
Hamiltonian and, by matching the excitation energies of the two systems, 
we can directly estimate the Landau parameters of the $t$$-$$J$ model.\\
Let us first note that at half-filling the EMD for the $t$$-$$J$ model is 
$n({\bf k})$$=$$1/2$, i.e., independent of momentum, and that this is similar 
to a band insulator where we would have $n({\bf k})$$=$$1$.  Also, the EMD 
for a state with one hole at momentum ${\bf k}_0$ and total spin $\sigma_0$ 
would be 
$n_\sigma({\bf k})$$=$$1$$-$$\delta_{{\bf k},-{\bf k}_0} \cdot 
\delta_{\sigma,\bar{\sigma}_0}$ 
in a band insulator, i.e., a constant with a `dip' at $-$${\bf k}_0$ and 
$\bar{\sigma}_0$.  The true single-hole EMD for the $t$$-$$J$ model however 
is quite different as we can see in Fig.~\ref{fig1}.  To emphasize small 
differences, this 
\noindent
figure shows $n_{\sigma}({\bf k})$ with its mean value $N_\sigma/N$ 
being subtracted (where $N_\sigma$ denotes the number of $\sigma$-spin 
electrons and $N$ is the cluster size).  We then find that the dip 
expected at $-{\bf k}_0$ and $\bar{\sigma}_0$ does indeed exist but 
that it is quite shallow.  Moreover, for both spin directions 
$n_{\sigma}({\bf k})$ acquires an additional, apparently smooth 
component, and the oscillation of this component around the mean value 
is almost independent of $\sigma$.  Another surprising result can be 
obtained by plotting the maximum amplitude of the EMD, i.e., 
$\Delta$$=$$n(0,0)$$-$$n(\pi,\pi)$ as a function of $\delta$ (see 
Fig.~\ref{fig2}).  We find that up to $\delta$ as large as $0.2$ the 
relation $\Delta\propto \delta$ holds with high accuracy, which is on 
one hand an unexpected result if we adopt a free-electron--like picture 
where a change of the electron density should affect $n({\bf k})$ 
predominantly near the noninteracting Fermi surface (which is far from 
either $(0,0)$ or $(\pi,\pi)$).  We could on the other hand understand 
this result if we were to assume that each hole `brings with it' the 
same smooth contribution as for the single hole (see Fig.~\ref{fig1}), 
so that for each hole number $n({\bf k})$ contains this component 
times the number of holes.  Such an assumption in turn is very 
natural\cite{EderBecker} if we assume that the doped holes are 
\noindent 
`spin bags'\cite{Schrieffer} where 
the bare hole oscillates rapidly inside a region of reduced spin 
correlation.  In this picture the smooth component of $n({\bf k})$ comes 
from the (incoherent and high-energy) oscillation of the hole `inside' 
the spin bag\cite{EderBecker}.  If this high-energy motion is 
sufficiently decoupled from much slower drift motion of the entire 
quasihole (i.e., the hole plus its dressing region), one may expect 
that this high-energy oscillation contributes identically to $n({\bf k})$ 
for each quasihole, whence the scaling of $n({\bf k})$ with hole number 
follows naturally.  
In `spin-charge-separation language' one might say that the physical 
quasiparticle is a firmly bound state of spinon and holon.  The rapid 
oscillation of the bound holon then might correspond to the motion of 
a boson in a localized orbital, and it is easy to understand that bosons 
which populate a localized orbital which is smaller than the cluster 
size have a momentum distribution proportional to the boson (=hole) density.  
Finally, one may expect that the picture proposed by Lee 
{\it et al.}\cite{LeeKimLee}, who modeled the charge degrees of freedom 
by bosons diffusing in a fluctuating magnetic field, would also give a 
diffuse component in $n({\bf k})$ proportional to the hole number.  
\\
With these ideas in mind we now assume that the smooth component in 
$n({\bf k})$ indeed accurately scales with hole number throughout 
the Brillouin zone; in other words, we write the EMD for a single hole: 
\begin{equation}
n_{\sigma}({\bf k}) = \frac{N_{\sigma}}{N} - 
Z \, \delta_{{\bf k},-{\bf k}_0}\cdot \delta_{\sigma,\bar{\sigma}_0} +
n_{\rm inc}({\bf k}).
\end{equation}
Then, in the first step, we `close the dip' at $-{\bf k}_0$ by replacing 
this $n({\bf k})$ value by that for a symmetry-equivalent ${\bf k}$-point; 
e.g., in Fig.~\ref{fig1} where ${\bf k}_0$$=$$(\pi/5,3\pi/5)$, we replace 
$n_\downarrow(-\pi/5,-3\pi/5) \rightarrow n_\downarrow(-3\pi/5,\pi/5)$.  
Subtracting now the constant term $N_\sigma/N$, we obtain the incoherent 
contribution $n_{\rm inc}({\bf k})$, which, as discussed above, should 
correspond to the oscillation of the bare hole inside the spin-bag--like 
quasiparticle.  For simplicity we average $n_{\rm inc}({\bf k})$ over spin 
directions and point-group operations (see Fig.~\ref{fig1}c).  
Next, we assume that also at higher doping levels each hole gives a two-fold 
contribution to the EMD: i.e., a `dip', which marks the total momentum of 
the {\it quasiparticle}, and the same $n_{\rm inc}({\bf k})$ as for the single 
hole.  If this is correct, the EMD for an eigenstate $|\Psi_\nu\rangle$ 
with $N_h$ holes (which we assume to be even) should take the form 
\begin{equation}
n({\bf k}) = \frac{N-N_h}{2} - Z \cdot n_{\rm coh}({\bf k}) +
N_h \cdot n_{\rm inc}({\bf k}).  
\label{2hole}
\end{equation}
Thereby the coherent part, $n_{\rm coh}({\bf k})$, corresponds to the EMD 
of $N_h$ spin-1/2 fermions which form a state with the same total momentum, 
spin, and point-group symmetry as $|\Psi_{\nu}\rangle$.  It can be readily 
obtained numerically by subtracting 
$\frac{N-N_h}{2}+N_h\cdot n_{\rm inc}({\bf k})$ 
from the exact EMD and dividing by $-Z$.  To judge the outcome of this 
procedure we now have to know how the $N_h$ quasiholes distribute themselves 
in ${\bf k}$-space for given total momentum and spin.  In other words, 
we need to guess the quasiparticle Hamiltonian.  The generic form is 
\begin{equation}
H = \sum_{ij\sigma} t_{ij} c_{i\sigma}^\dagger
c_{j\sigma}
+ \sum_{ij} ( V_{ij} n_i n_j + J_{ij} {\bf S}_i \cdot {\bf S}_j ).
\label{heff}
\end{equation}
We have chosen a next-nearest-neighbor hopping dispersion, i.e., only the 
hopping integrals $t_{11}$ and $t_{20}$ between the second-nearest 
(i.e., (1,1)) and third-nearest (i.e., (2,0)) neighbors are different 
from zero.  Moreover we retain nearest-neighbor density-density and 
exchange interactions, $V_{10}$ and $J_{10}$.  In view of the many 
complicated ways in which two `dressed holes' in an antiferromagnet can
interact with each other\cite{Interaction}, this may be a very oversimplified 
choice; however this will turn out to be sufficient to get reasonable 
results\cite{afvh}.  The EMD for the different eigenstates of this 
Hamiltonian are easily computed and can then be compared to the 
`coherent components' obtained by the above subtraction procedure for 
the $t$$-$$J$ model.\\  
We should note, before discussing the results, that the procedure is 
based on a whole sequence of quite strong assumptions: 
we assume that $n({\bf k})$ really can be meaningfully decomposed into 
two components, as in Eq.~(\ref{2hole}), that the smooth component indeed 
scales accurately with the hole number, and consequently is the same for 
all low-energy states of the $t$$-$$J$ model, and finally that a one-to-one 
correspondence exists between the states of the $t$$-$$J$ model and those 
of the quasihole Hamiltonian.  
The outcome of this procedure, which is shown in Figs.~\ref{fig3} and 
\ref{fig4}, is however all the more surprising.  To begin with, it is 
obvious that the calculated `coherent components' $n_{\rm coh}({\bf k})$ 
are indeed coherent, i.e., in most cases they are sharply localized in 
momentum space and nearly zero almost everywhere in the Brillouin zone.  
This shows that for most ${\bf k}$-points our scaling procedure indeed gives 
a good guess for $n({\bf k})$.  We would like to stress again that we have 
subtracted one and the same estimated incoherent contribution from the 
exact EMD for {\it all} the different low-energy states in the $t$$-$$J$ 
model to obtain the $n_{\rm coh}({\bf k})$.  
Next, comparing $n_{\rm coh}({\bf k})$ with the EMD for the model system, 
a remarkable correspondence is obvious.  Under $20$ different states 
shown in Figs.~\ref{fig3} and \ref{fig4} only the singlet state with 
${\bf k}$$=$$(\pi/3,\pi/3)$ in the $18$-site cluster deviates significantly.  
It should be noted that, with the exception of the high-symmetry points 
$(0,0)$ and $(\pi,\pi)$, Figures \ref{fig3} and \ref{fig4} comprise the 
lowest singlet and triplet 
\noindent
states for all the allowed momenta in both $18$- 
and $20$-site clusters; a similarly good correspondence can also be seen for 
the $16$-site cluster, which we do not show here.  The reason for the 
deviations in the state with ${\bf k}$$=$$(\pi/3,\pi/3)$ and $S$$=$$0$ are 
unclear; this state may correspond to a charge-density-wave or stripe-like 
hole arrangement, as can be seen for larger values of $J$\cite{TK}.  
Finally, the correspondence between the $t$$-$$J$ model and the quasiparticle 
system also becomes apparent in a comparison of the excitation energies 
in Fig.~\ref{fig5}.  
While we may not expect to obtain perfect agreement 
with our rather simplified choice of the interaction terms in Eq.~(\ref{heff}), 
the relative errors are mostly $\leq 20$\% and the `dispersion' of the 
excitation energy is reproduced reasonably well.  The optimized parameter 
values in Eq.~(\ref{heff}) are $t_{11}$$=$$0.255$, $t_{20}$$=$$0.15$, 
\noindent 
$V_{10}$$=-$$0.6375$, and $J_{10}=0.15$.
\\  
In summary, we have shown that the EMD for all the low-energy states 
of the small-cluster $t$$-$$J$ model in 2D can be accurately decomposed into 
an incoherent contribution, which has a boson-like dependence on hole density 
and is independent of the total momentum and spin of the respective state, 
\noindent
and a coherent contribution, which corresponds to that of spin-1/2 quasiholes.  
Using the coherent contributions we have directly established the Landau 
mapping between the low-energy states of the $t$$-$$J$ model and those of 
a `quasiparticle Hamiltonian' describing the system as a Fermi liquid of 
spin-1/2 particles corresponding to the doped holes.  The quasiparticle 
dispersion is dominated by the next-nearest-neighbor hopping, as would be 
expected for a dispersion dominated by antiferromagnetic spin 
correlation\cite{dispersionI}, but also for a `spinon' in an RVB 
state\cite{dispersionII}.  
The interaction between quasiparticles is 
dominated by a nearest-neighbor attraction.  Our results establish, in our 
opinion conclusively, that the small clusters behave like the finite-size 
equivalents of a Landau Fermi-liquid, in that there is a one-to-one mapping 
between the exact eigenstates of the interacting system and those of an 
interacting system of quasiparticles.  The most obvious extrapolation to 
the infinite systems is of course that the same quasiparticle Hamiltonian 
also describes the thermodynamic limit.  There is good reason to 
believe that the cluster diagonalization quite accurately describes short-range 
processes, which are the dominant ones in strongly correlated systems; 
for the paramagnetic regime with its very short-range spin correlations it 
seems moreover plausible that longer-range processes are of little 
importance for either propagation or interaction of the holes, so that we 
believe that the extrapolation of the quasiparticle Hamiltonian from the 
clusters to the infinite system is a quite reasonable approximation.  
Unless the interaction between holes (which may be either the interaction 
intrinsic to the $t$$-$$J$ model or the extra Coulomb repulsion) drives the 
system into a charge-density-wave state\cite{TK} (as may be the case in 
La$_{1.875}$Sr$_{0.125}$CuO$_4$), the system should thus behave like a 
Fermi liquid with a hole-pocket Fermi surface centered on 
${\bf k}$$=$$(\pi/2,\pi/2)$.\\ 
This work was supported in part by Grant-in-Aid for Scientific 
Research from the Ministry of Education, Science, and Culture of 
Japan.  Financial supports of S.~N. by Sasakawa Scientific Research 
Grant from the Japan Science Society and of Y.~O. by Saneyoshi Foundation 
are gratefully acknowledged.  
Computations were carried out in Computer Centers of the University of 
Groningen, the Institute for Solid State Physics, University of Tokyo, 
and the Institute for Molecular Science, Okazaki National Research 
Organization.  
 
\end{multicols}
\end{document}